\documentclass[amsmath, amssymb, reprint, aps, onecolumn]{revtex4-2}

\usepackage{orcidlink}

\newcommand{\SSD}{Sensor Science Division, National Institute of Standards and Technology, Gaithersburg, Maryland 20899, USA}
\newcommand{\CTL}{Communications Technology Laboratory, National Institute of Standards and Technology, Boulder, Colorado 80305, USA}

\begin{document}

\title{Atomic and molecular systems for radiation thermometry}

\author{Stephen P. Eckel\orcidlink{0000-0002-8887-0320}}
\email[Corresponding author: ]{stephen.eckel@nist.gov}
\affiliation{\SSD}
\author{Christopher L. Holloway\orcidlink{0000-0002-4592-9935}}
\affiliation{\CTL}
\author{Eric B. Norrgard\orcidlink{0000-0002-8715-4648}}
\affiliation{\SSD}
\author{Nikunjkumar Prajapati\orcidlink{0000-0002-7779-9741}}
\affiliation{\CTL}
\author{Noah~Schlossberger\orcidlink{0000-0001-9573-8152}}
\affiliation{\CTL}
\author{Matthew Simons\orcidlink{0000-0001-9418-7520}}
\affiliation{\CTL}




\begin{abstract}
Atoms and simple molecules are excellent candidates for new standards and sensors because they are both all identical and their properties are determined by the immutable laws of quantum physics. Here, we introduce the concept of building a standard and sensor of radiative temperature using atoms and molecules. Such standards are based on precise measurement of the rate at which blackbody radiation (BBR) either excites or stimulates emission for a given atomic transition. We summarize the recent results of two experiments while detailing the rate equation models required for their interpretation. The cold atom thermometer (CAT) uses a gas of laser cooled $^{85}$Rb Rydberg atoms to probe the BBR spectrum near 130~GHz. This primary, {\it i.e.}, not traceable to a measurement of like kind, temperature measurement currently has a total uncertainty of approximately 1~\%, with clear paths toward improvement.
The compact blackbody radiation atomic sensor (CoBRAS) uses a vapour of $^{85}$Rb and monitors fluorescence from states that are either populated by BBR or populated by spontaneous emission to measure the blackbody spectrum near 24.5~THz.
The CoBRAS has an excellent relative precision of $u(T)\approx 0.13$~K, with a clear path toward implementing a primary measurement.
\end{abstract}



\maketitle

\section{Introduction}
From the days of John Dalton, the scientific community has realized that all atoms of a certain species and isotope are identical: they share the same mass, energy levels, polarizability, etc.
These properties are dictated by the immutable laws of atomic physics.
Because of the constancy and consistency of their properties, atoms and, by extension, simple molecules make excellent potential measurement sensors and standards.
To date, atoms have been used as fundamental standards of time~\cite{ludlow_optical_2015}, and have been used as the basis for magnetic field~\cite{kitching_chip-scale_2018}, electric field~\cite{schlossberger_rydberg_2024}, vacuum pressure~\cite{barker_precise_2022}, and inertial~\cite{cronin_optics_2009} sensors.

Over the past several years, we have been striving to use atoms and simple molecules as standards and sensors for radiative temperature.
As realized by Einstein~\cite{Einstein1916}, blackbody radiation at a temperature $T$ can drive transitions between quantum states, both by absorbing photons and by stimulating emission of photons.
The rate of absorption or stimulated emission between two states $i$ and $j$, assuming no degeneracy, is the product of two quantities.
The first is the photon energy density per unit angular frequency, $U_\omega(\omega, T)$, which is a function of both the angular frequency $\omega$ and $T$.
The second is the dipole transition dipole matrix element squared, $\left|\left<i|d|j\right>\right|^2$~\cite{hilborn_einstein_1982}.
Measured in s$^{-1}$, the transition rate is
\begin{align}
\Omega_{ij} & = \frac{\pi}{3\epsilon_0 \hbar^2}\left|\left<i|d|j\right>\right|^2 U_\omega(\omega_{ij}, T) \\& = \frac{\omega_{ij}^3}{3\epsilon_0 \hbar\pi c^3}\left|\left<i|d|j\right>\right|^2 \frac{1}{e^{\hbar\omega_{ij}/k_B T}-1}\,.\label{eq:transition_rate}
\end{align}
Appearing in \eqref{eq:transition_rate} are three defining constants of the SI---the speed of light $c$, the reduced Planck constant $\hbar$, and the Boltzmann constant $k_{\rm B}$---and one derived physical constant---the permittivity of free space $\epsilon_0$.
Also appearing in \eqref{eq:transition_rate} are two atomic or molecular properties: the transition frequency $\omega_{ij} = |E_j-E_i|/\hbar$, with $E_{i,j}$ as the energy of the $i,j$th state and the dipole matrix element $|\left<i|d|j\right>|^2$.
Thus, \eqref{eq:transition_rate} relates transition rates to temperature through fundamental constants and atomic properties, making population transfer an ideal candidate for an atomic temperature sensor.
Moreover, because such a sensor need not be traceable to a measurement of like kind ({\it in casu}, a different thermometer), it can also serve as a primary standard.

\begin{figure*}
    \centering
    \includegraphics[width = \linewidth]{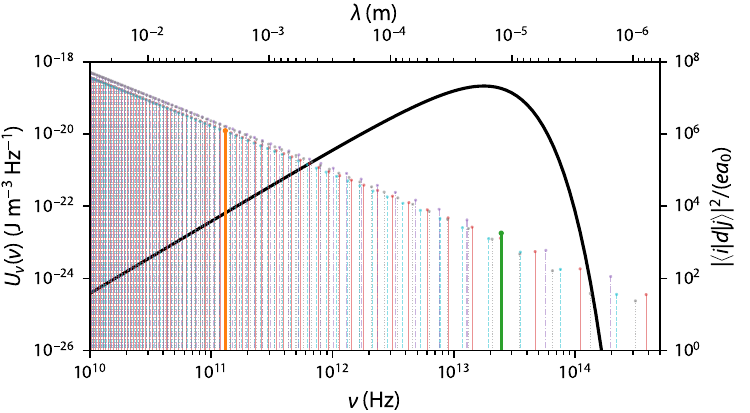}
    \caption{
    Planck's law at $T=300$~K (black curve, left scale) and a stick spectrum showing different atomic dipole strengths (right scale) for different transitions in Rb vs. frequency (bottom axis) and wavelength (top axis).
    Transitions between $n{\rm S}_{1/2}\rightarrow n{\rm P}_{3/2}$ are show as red, solid sticks; $n{\rm S}_{1/2}\rightarrow (n-1){\rm P}_{3/2}$ are dashed, cyan sticks; $n{\rm P}_{3/2}\rightarrow (n-1){\rm D}_{5/2}$ are dashed-dot purple sticks; and $n{\rm P}_{3/2}\rightarrow (n-2){\rm D}_{5/2}$ are dotted, gray sticks. 
    The highlighted orange and green sticks correspond to the transitions used in the experiments of Secs.~\ref{sec:rydberg_thermometry} and~\ref{sec:fluorescence_ratio}, respectively.
    }
    \label{fig:transitions_and_BBR}
\end{figure*}

For almost any conceivable $\omega$ of interest, at least one transition can be found in any given atom or molecule with $\omega_{ij}\approx \omega$. 
Indeed, even in a simple alkali atom like Rb, transitions can be identified across the full Planck spectrum.
Figure~\ref{fig:transitions_and_BBR} shows a subset of transitions in Rb taken from a calculation using the python package ARC~\cite{sibalic_arc_2017} and, when available, the NIST spectral database~\cite{kramida_nist_2024} together with $U_\omega(\omega, T=300~\mbox{K})$.
Specifically, we show transitions between $n{\rm S}\rightarrow (n,n+1){\rm P}$ states and $n{\rm P}\rightarrow (n-2,n-1){\rm D}$ states, where $n$ is the principal quantum number.
Transitions can be observed over the entire infrared regime and into the microwave regime.
Selecting the band of interest for thermometry can thus be as simple as selecting a different transition within the atom.
Transitions become denser at lower frequency, corresponding to increasing $n$.

Moreover, each transition acts as a natural spectral filter, set by the natural lifetime of the state.
Such lifetimes tend to be between a few tens of nanoseconds for transitions at $10^{14}$~THz and several hundred microseconds in the microwave regime, corresponding to effective filter bandwidths of roughly 10~MHz to less than 1~kHz.

To understand the sensitivity of such an atomic thermometer at frequency $\omega$, we divide $U_\omega(\omega)$ into two regimes: the regime where $\hbar\omega/k_B T\ll 1$ and $U_\omega(\omega) \propto \omega^2$, and the regime where $\hbar\omega/k_B T\gg 1$ and $U_\omega(\omega)\propto \exp(-\hbar\omega/k_B T)$.
In the former regime, an interesting cancellation occurs: $\left|\left<i|d|j\right>\right|^2\propto \omega^{-2}$ and $U_\omega(\omega)\propto\omega^2$, and thus $\Omega_{ij}$ is roughly constant.
Thus, in this regime, the atomic thermometer's precision is independent of frequency or, alternatively, of chosen internal transition.
Indeed, Ref.~\cite{norrgard_quantum_2021} calculates the anticipated sensitivity per atom in this regime, and found it to be roughly independent of $\omega_{ij}$ and $u(T)/T\approx 2/\sqrt{N}$, where $N$ is the number of atoms probed in the experiment.
Measuring approximately $10^{10}$ atoms would then yield a precision of $u(T)/T\approx 10^{-5}$~\cite{norrgard_quantum_2021}.
In the latter regime, the situation is more complicated as the dependence of $U_\omega(\omega)$ transitions to exponential, while $\left|\left<i|d|j\right>\right|^2$ continues along its $\omega^{-2}$ dependence.
Nevertheless, the exponential dependence of $U_\omega(\omega)$ on $T$ will ensure that the signal is exponential in temperature. 

The advantages of being a primary sensor, frequency selection, narrow spectral width, and good sensitivity are counterbalanced by two disadvantages.
First, measuring $\Omega_{ij}$ requires measuring the populations of the states in an ensemble of atoms over time.
Such a measurement can prove difficult, especially if the two states involved in the transition of interest are both excited states.
In this case, there are transitions to other states, particularly lower states that complicate the dynamics.
Here, a rate equation model, introduced in Sec.~\ref{sec:rate_eqns}, needs to be constructed to capture the full behavior of the multi-level system.
Additional uncertainties due to inaccurate parameters of the rate equation model then become relevant.

Second, the accuracy of such a sensor is currently limited by atomic theory.
While the frequencies $\omega_{ij}$ can be independently measured with better than 1 part in $10^8$ accuracy or better, the knowledge of atomic dipole matrix elements $|\left<i|d|j\right>|^2$ is significantly less accurate.
Again, taking the Rb atom as our example, some of the best known transitions in the near IR (around 10~$\mu$m or shorter) have roughly 1~\% uncertainty~\cite{safronova_critically_2011, barakhshan_portal_2022}.
For transitions in the microwave regime, which correspond to transitions within the Rydberg manifold or, equivalently, transitions between states with high principal quantum number $n$, the uncertainty can be less than 1~\%~\cite{sibalic_arc_2017}.

In Sections~\ref{sec:rydberg_thermometry} and~\ref{sec:fluorescence_ratio}, we examine two recently published examples of radiation thermometry~\cite{schlossberger_primary_2025, mantia_compact_2024} with special emphasis on the rate equation model of Sec.~\ref{sec:rate_eqns}.
These two thermometers both use Rb atoms, yet they probe vastly different sections of the blackbody spectrum, as shown in Fig.~\ref{fig:transitions_and_BBR}.
Fluorescence ratio thermometry, discussed in Section~\ref{sec:fluorescence_ratio}, probes the blackbody spectrum at 12.2~$\mu$m (24.5~THz) by examining the  $7{\rm S}\rightarrow 6{\rm D}$ transition.
Rydberg state selective ionization thermometry, discussed in Section~\ref{sec:rydberg_thermometry}, probes the blackbody spectrum at 2.3~mm (130~GHz) by examining the  $32{\rm S}\rightarrow 32{\rm P}$ transition.
Finally, we will conclude by discussing potential avenues for improvement and previewing a couple possible other future measurements for quantum-based blackbody radiation thermometry.

\section{The rate equations}
\label{sec:rate_eqns}
In general, we model the internal state populations $n_i$, where $i=1,2,\cdots, N$, via a set of coupled rate equations,
\begin{equation}
    \frac{\partial}{\partial t} \begin{bmatrix} n_1 \\ n_2 \\ \vdots \\ n_N \end{bmatrix} = 
    \begin{bmatrix}
   -\Omega_1 -\Gamma_1 & \Omega_{12} + \Gamma_{21} & \cdots & \Omega_{1N} + \Gamma_{N1} \\
    \Omega_{12} & -\Omega_2-\Gamma_2 & \cdots & \Omega_{2N}+\Gamma_{N2} \\
    \vdots & \vdots & \ddots & \vdots \\
    \Omega_{1N} & \Omega_{2N} & \cdots & -\Omega_N-\Gamma_N
    \end{bmatrix}
    \begin{bmatrix} n_1 \\ n_2 \\ \vdots \\ n_N  \end{bmatrix}\label{eq:rate_equations},
\end{equation}
where $\Omega_{ij}$ is given by \eqref{eq:transition_rate}, 
\begin{equation}
    \Gamma_{ij} = \frac{\omega_{ij}^3}{3 \epsilon_0 \hbar \pi c^3}\left|\left<i|d|j\right>\right|^2 = \Omega_{ij}(e^{\hbar\omega_{ij}/k_B T}-1)
\end{equation}
is the fraction of the spontaneous decay from state $i$ that falls into state $j$, $\Gamma_i = \sum_j\Gamma_{ij}$ is the total spontaneous decay rate of state $i$, and $\Omega_i = \sum_j\Omega_{ij}$ is the total stimulated rate of state $i$.
Note that, by construction, the states are energy ordered.
While the matrix in \eqref{eq:rate_equations} does not depend on time, it does depend on temperature through the $\Omega_{ij}$.
At first glance, extracting $T$ from \eqref{eq:rate_equations} might appear hopeless, given the large number of populations that would have to be measured ({\it e.g.} $n_1$, $n_2$, $\cdots$, $n_N$) and the large number of atomic physics parameters that must be determined either theoretically or experimentally.
However, \eqref{eq:rate_equations} can generally be greatly simplified and elements of it can be isolated by choosing the measurement scheme carefully, as we shall see.

\subsection{The ideal two level system}

For the ideal two-level system, \eqref{eq:rate_equations} greatly simplifies.  In that case, we take $\Omega_{12}=\Omega$ and $\Gamma_1=0$ and $\Gamma_2=\Gamma$ and find
\begin{equation}
    \frac{\partial}{\partial t} \begin{bmatrix} n_1 \\ n_2 \end{bmatrix} = 
    \begin{bmatrix}
    -\Omega & \Omega+\Gamma \\
    \Omega & -\Gamma-\Omega 
    \end{bmatrix}
    \begin{bmatrix} n_1 \\ n_2 \end{bmatrix}\label{eq:two_state_eate_eq},
\end{equation}
This has the solution 
\begin{align}
    n_0 (t) & = \frac{\Gamma+\Omega+\Omega e^{-t(\Gamma+2\Omega)}}{\Gamma+2\Omega} \\
    n_1 (t) & = \frac{\Omega(1- e^{-t(\Gamma+2\Omega)})}{\Gamma+2\Omega}.
\end{align}
Note that in the limit that $t\rightarrow \infty$, 
\begin{align}
    \label{eq:thermal_equlibrium}
    n_1 = \frac{\Omega}{\Gamma+2\Omega} = \frac{e^{-\hbar\omega/k_{\rm B} T}}{1+ e^{-\hbar\omega/k_{\rm B} T}}\,.
\end{align}
showing that the two states tend toward thermal equilibrium, as established by the Boltzmann distribution.
This fact was also realized by Einstein in his seminal paper~\cite{Einstein1916}.

\subsection{The potential of laser-cooled molecules}

Molecules offer a large hierarchy of potential states, with rotational, vibrational, and electronic degrees of freedom.
In Ref.~\cite{norrgard_quantum_2021}, the authors proposed using the two lowest vibrational states ($v=0,1$)in a laser-coolable molecule to realize the two level system described above.
In order to prevent population of $v\geq2$ states, the authors proposed using a combination of lasers and microwaves to effectively repump these states into the two-level system.
Provided that the pumping rates are fast compared to $\Omega$ and $\Gamma$, which are on the order of 1~s$^{-1}$, the system can be effectively closed into a two-level system.
As a two level system, measurement of the exponential trend toward equilibrium, $n_1(t)$ and $n_2(t)$ and the equilibrium values, $n_1(t\rightarrow\infty)$ and $n_2(t\rightarrow\infty)$ can fully characterize the system, measuring both $T$ and $\left|\left<i|d|j\right>\right|^2$~\cite{norrgard_quantum_2021}.

Subsequently, the authors of Ref.~\cite{vilas_blackbody_2023} measured the evolution of laser-cooled CaOH molecules under the influence of BBR.
Without the repumping scheme of Ref.~\cite{norrgard_quantum_2021}, the molecules were allowed to evolve freely up the more complicated vibrational levels of a linear, triatomic molecule.
Through a rate equation model, the authors of Ref.~\cite{vilas_blackbody_2023} extracted key parameters of the molecule, including the dipole moment and the vacuum lifetime of the vibrational states.
These experimental values were compared to {\it ab initio} calculations and were found to be in good agreement.
While not a thermometer, Ref.~\cite{vilas_blackbody_2023} shows the power of using a thermal blackbody environment together with a rate equation model to extract key atomic and molecular parameters with good accuracy. 

\section{Rydberg state selective ionization thermometry}
\label{sec:rydberg_thermometry}
We begin our discussion of atom- and state-transfer-based thermometry with the cold atom thermometer (CAT), which was recently published in detail in Ref.~\cite{schlossberger_primary_2025}.
Here, we present only the salient details.
The CAT builds upon early work on BBR-induced state transfer~\cite{gallagher_interactions_1979, hildebrandt_interaction_1981, galvez_multistep_1995}. 

\begin{figure}
    \centering
    \includegraphics[width=\columnwidth]{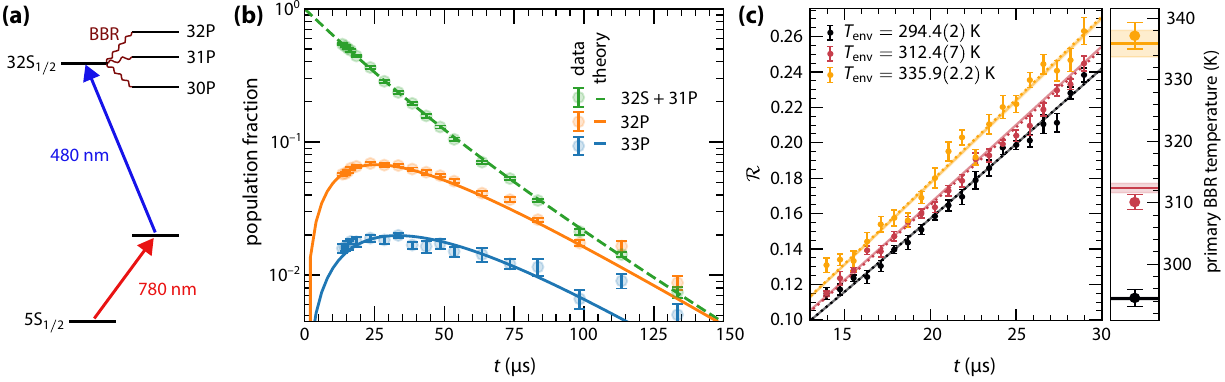}
    \caption{
    (a) Relevant level diagram of $^{85}$Rb for the cold atom thermometer (CAT) experiment.
    Atoms in the $5{\rm S}_{1/2}$ ground state are pumped using a 780~nm laser into $5{\rm P}_{3/2}$ intermediate state and further excited using a 480~nm laser into the $32{\rm S}_{1/2}$ state.
    Blackbody radiation (BBR) then transfers some of the atoms into nearby P states, including 30P, 31P, and 32P.
    (b) Relative population in various states as a function of BBR interaction time $t$ at $T=296$~K.
    The theoretical curves are calculated by a rate equation model \eqref{eq:rate_equations}, using decay rates from ARC~\cite{sibalic_arc_2017}.
    (c, left panel) Ratio of the 32S+31P peak to the 32P peak $\mathcal{R}$ as a function of blackbody interaction time.
    Data are shown as points, theoretical prediction with two fit parameters shown as solid curves.
    (c, right panel) The atomic-determined temperatures (points) together with the contact thermometry determined temperatures (horizontal lines with corresponding uncertainty bars).
    Figure adapted from Ref.~\cite{schlossberger_primary_2025}.
    }
    \label{fig:cat_experiment}
\end{figure}
The relevant level structure is shown in Fig.~\ref{fig:cat_experiment}a. Two linearly polarized lasers drive the $5{\rm S}_{1/2}(F'=3)\rightarrow 5{\rm P}_{3/2}(F'=4)$ transition at 780.1242~nm and the $5{\rm P}_{3/2}(F'=4)\rightarrow 32{\rm S}_{1/2}$ transition. (The hyperfine states in the excited $32{\rm S}_{1/2}$ state are unresolved.)
From there, blackbody radiation drives transitions to nearby P states, including $31{\rm P}$, $32{\rm P}$, etc.
Once these states have appreciable population, BBR can further drive transitions from ${\rm P}\rightarrow {\rm S}$ or ${\rm P}\rightarrow{\rm D}$, and so forth.
However, the dominant transition in this process is between the initially populated $32{\rm S}\rightarrow 32{\rm P}$, a transition driven by the 130.507~GHz component of the microwave BBR spectrum.

The states involved in this system are of high principal quantum number and are known as Rydberg states.
Collisions between atoms excited to the Rydberg state, heretofore referred to as Rydberg atoms, and ground-state atoms can cause the nearly-free electron of the Rydberg atom to change state or be completely ionized~\cite{gallagher_rydberg_1994}.
To suppress these effects, we laser cool a $^{85}$Rb cloud with approximately $2\times10^6$ atoms to temperatures of $\sim 0.5$~mK~\cite{schlossberger_primary_2025}.
(We note we did not optimize our MOT for achieving the typical Doppler-limited temperature of roughly 150~$\mu$K.)

The additional complication of laser cooling makes the experiment step-wise.
First, a cold atomic cloud is prepared and held briefly in a three-dimensional magneto-optical trap.
The cloud is then released by turning off the trap and subsequently expands due to its non-zero temperature and falls under the influence of gravity.
While falling, the two lasers at 780~nm and 480~nm are pulsed using acoustic-optic modulators for roughly 3~$\mu$s to pump roughly $5\times10^3$ atoms into the $32{\rm S}_{1/2}$ state.
This pumping process marks $t=0$.
The atoms then evolve under the influence of blackbody radiation for a time $t$.

We measure the population transfer between states using selective field ionization.
In the process, an electric field is ramped from zero volts to of the order of 80~kV/m in roughly 10~$\mu$s.
The electric field tilts the potential that binds the electron to the nucleus, causing the electron to effectively spill out.
At what field the electron spills out is dependent on its initial state, and thus the timing of the ionization with respect to the electric field ramp yields information regarding the electron's initial state.
Ionization peaks from closely spaced levels are not necessarily resolved.
For this reason, our experiment cannot distinguish the closely spaced 31P from 32S, but we can clearly distinguish the more energy separated 32P and 33P from 32S.
The procedure for identifying the ionization peaks can be found in Ref.~\cite{schlossberger_primary_2025}.

For $t>0$, the evolution again follows the rate equations~\eqref{eq:rate_equations}.
To gain insight, let us consider a simplified version of \eqref{eq:rate_equations} that contains just 32S and 32P,
\begin{equation}
\frac{\partial}{\partial t} \begin{bmatrix} n_{32{\rm S}} \\ n_{32{\rm P}} \end{bmatrix} = 
    \begin{bmatrix}
    -\Gamma_{32{\rm S}}-\Omega_{32{\rm S},32{\rm P}} & \Omega_{32{\rm S},32{\rm P}} \\
    \Omega_{32{\rm S},32{\rm P}}  & -\Gamma_{32{\rm P}} -\Omega_{32{\rm S},32{\rm P}}  
    \end{bmatrix}
    \begin{bmatrix} n_{32{\rm S}} \\ n_{32{\rm P}}\end{bmatrix}\,.
    \label{eq:cat_rate_equations}
\end{equation}
The solution given $n_{32{\rm S}}(t=0)=1$ and $n_{32{\rm P}}(t=0)=0$ is
\begin{align}
    n_{\rm 32S}(t) & = \frac{1}{2\Omega'}e^{-\bar{\Gamma}t}\left(e^{- \Omega_+ t}[\Gamma_{32{\rm S}} - \Gamma_{32{\rm P}} +\Omega']-e^{- \Omega_- t}[\Gamma_{32{\rm S}} - \Gamma_{32{\rm P}} -\Omega']\right)\\
    n_{\rm 32P}(t) & = \frac{\Omega_{32{\rm S},32{\rm P}} }{2\Omega'}e^{-\bar{\Gamma}t}\left(e^{-\Omega_+t} -e^{- \Omega_- t}\right)
\end{align}
with $2\Omega' = \sqrt{(\Delta\Gamma/2)^2 + \Omega_{32{\rm S},32{\rm P}}}$, $\Omega_+=\Omega_{32{\rm S},32{\rm P}} + \Omega'$, $\Omega_-=\Omega_{32{\rm S},32{\rm P}} - \Omega'$, $\Delta\Gamma = \Gamma_{32{\rm P}}-\Gamma_{32{\rm S}}$, and $\bar{\Gamma} = (\Gamma_{32{\rm P}}+\Gamma_{32{\rm S}})/2$.

Several features of the solution are immediately apparent.
At short times, when $\Omega_+t\ll 1$ and $\Omega_-\ll 1$, $n_{\rm 31P}\propto \Omega_{\rm 32S,31P} t$.
Because $\Omega_{\rm 32S,31P}$ is itself proportional to $u_\omega$, the initial rate of population growth of 31P is closely tied to $T$.
At long times, when $\Omega't\gg1$ and $\Omega_{32{\rm S},32{\rm P}}t\gg1$, then the both $n_{32{\rm S}}$ and $n_{31{\rm S}}$ tend toward the same exponential behavior, with $n_{\rm 32S},n_{\rm 32P}\propto e^{-\bar{\Gamma} t}$.
Physically, the fact that the two states decay at the average of the two rates is related to the fact that the two states are in ``equilibrium'' with each other for $\Omega_{32{\rm S},32{\rm P}}t\gg1$.
 
These features are immediately evident in the CAT data.
Fig.~\ref{fig:cat_experiment}b shows the  evolution of the relative populations for $T=296$~K.
The population begins in 32S, and the population in that state and the nearby, unresolved 31P begin to decay away nearly exponentially with time.
(There is also some transfer into 31P, but that transfer cannot be easily resolved from the large population in 32S.)
The populations in 32P and 33P both peak around $t=25$~$\mu$s, and then they also transition to exponential decay at long times.
The overall decay at long times is set by the lifetime of the states in the Rydberg manifold.
The temporal location and size of the peak are both related to $T$: as $T$ increases, the time of the peak decreases and the peak height increases.

\begin{table}[b]
    \centering
    \caption{\label{tab:cat_uncertainty} Relative uncertainty contributions to $T$ in the CAT.  All uncertainites are $k=1$. Adapted from Ref.~\cite{schlossberger_primary_2025}}
    \begin{tabular}{p{5cm} c}
    \hline\hline
        \textbf{source} & $\sigma_T/T$\\ \hline
        detector non-linearity & 0.002\\
        ion time-of-flight overlap & 0.005 \\
        determination of $t = 0$ & 0.002\\31
        detection signal artifacts & 0.003\\
        time-dependent gain calibration & $1\times 10^{-5}$\\
        \hline
        total, type B    & 0.006\\
        total, type A$^*$ & 0.005 \\
        \hline
        total & 0.008 \\
        \hline\hline
    \end{tabular}
    \vspace*{-4pt}
\end{table}

A more efficient way of quantifying the evolution is to track the ratio of the peak area of 32S+31P to 32P, $\mathcal{R}$.
This ratio is most sensitive to $\Omega_{32{\rm S}, 32{\rm P}}$
Figure~\ref{fig:cat_experiment}(c) shows this ratio, which increases almost linearly with $t$.
The slope of the ratio with time, $\dot{\mathcal{R}}$, is proportional to $T$, with $d\dot{\mathcal{R}}/dT\approx 27$~K$^{-1}$s$^{-1}$.
Three separate environmental temperatures are shown in Fig.~\ref{fig:cat_experiment}c, ranging from 294~K to 336~K.
The atomic temperatures from the slope $\dot{\mathcal{R}}$ are shown together with the expected BBR temperatures, the latter determined by contact thermometers that measure the temperature of the walls of the vacuum chamber establish BBR environment at the location of the atoms.
Agreement is observed, verifying the accuracy of the CAT.

Table~\ref{tab:cat_uncertainty} shows the uncertainty budget for the CAT, which assumes no uncertainty in $\Omega_{32{\rm S}, 32{\rm P}}$.
There are systematic effects due to the non-linearity of the avalanche-based ion detector, overlap of the 32P and 32S+31P peaks in the ion time-of-flight signal due to insufficient bandwidth, determination of $t=0$ given the non-zero amount of time required to pump the atoms into 32S, and other signal artifacts.
The three largest systematics may be improved by switching detectors: rather than using avalanche detectors, using Faraday cups with charge sensitive amplifers should offer sufficient dynamic range, linearity, and bandwidth to improve these systematics.
Improved optical pumping and faster selective field ionization ramps could further improve determination of $t=0$.
It is not inconceivable that with these few minor improvements the total type A uncertainty could drop below 0.1~\%.
At this level, it would be comparable to other blackbody sources and radiometers operating in the microwave regime.
Crucially, it will also be at the limits of the accuracy of atomic theory calculations.
We shall postpone that discussion for the conclusion.

\section{CoBRAS: Fluorescence ratio thermometry}
\label{sec:fluorescence_ratio}

To probe BBR at a wavelength closer to that at which the blackbody spectrum has its peak requires probing states at lower principal quantum number.
In this section, we shall discuss the compact blackbody radiation atomic sensor (CoBRAS), which probes $\Omega_{7P\rightarrow 6D}$.

\begin{figure}
    \centering
    \includegraphics{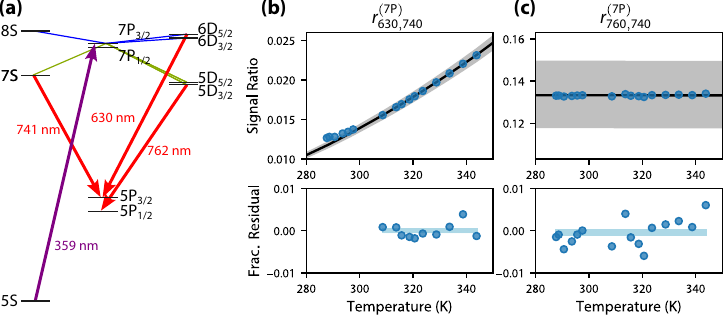}
    \caption{(a) Simplified level diagram for the CoBRAS.  A laser at 359~nm excites atoms from the ground 5S state to the $7{\rm P}_{3/2}$ state.
    Atoms can then be further excited by blackbody radiation into states including the 8S or 6D (blue lines) or decay via spontaneous emission to lower energy states including the 7S and 5D states (green lines).
    Fluorescence from 6D, 7S, and 5D to 5P$_{3/2}$ (red arrows) is monitored by PMTs with the appropriate filter wavelengths indicated.
    (b, top) The ratio of PMT signals recorded with the 630~nm filter to the 740~nm filter vs. temperature.
    Blue points indicate the data; solid black curve the thoeretical prediction with one fitted parameter.
    Gray shaded region indicates the uncertainty in the theory due to atomic physics parameters.
    (b, bottom) fractional residual of between the data and prediction.
    Blue band shows the mean and root-mean-squared uncertainty.
    (c) The ratio of PMT signals recorded with the 762~nm filter to the 740~nm filter (top) and the fractional residuals between the data and prediction (bottom) vs. temperature.
    Same symbols as is (b).
    Figure adapted from Ref.~\cite{mantia_compact_2024}.}
    \label{fig:cobras_experiment}
\end{figure}

Figure~\ref{fig:cobras_experiment}(a) shows the relevant energy diagram.  
A 359~nm laser drives atoms from the $5{\rm S}_{1/2}$ ground state of $^{85}$Rb to the $7{\rm P}_{3/2}$ excited state.
From there, atoms can be pumped into the 6D state by 12.2~$\mu$m (24.5~THz) BBR radiation.
Alternatively, the atoms can spontaneously decay to either the 7S or the 5D states, among others.
Once in a high-lying S or D state, the atoms will then further spontaneously decay, possibly to the $5{\rm P}_{3/2}$ state.
When decaying to 5P$_{3/2}$ from 7S, 5D, or 6D states, the atoms emit 741~nm, 762~nm, and 630~nm photons, respectively.
CoBRAS detects these fluorescence photons and uses ratios of their relative intensity to determine the temperature.

Thus, the CoBRAS is quite similar to other luminesence thermometry techniques~\cite{dramicanin_trends_2020,wang_review_2021}.
In particular, it bears resemblance to fluorescence ratio luminesence thermometry (FRLT)~\cite{HAN2019,Jia2020,wang_review_2021,Li2021}, which uses rare-earth ions embedded into a solid.
Because of the complicated nature of both the rare earth ions and the solid matrix, it appears unlikely that FRLT will be calculable from first principles.
Moreover, because the ions are doped into a solid, it is not necessarily clear that all sample realizations will be identical as they are for free atoms.

Experimentally, the 359~nm laser beam passes through the long axis of a cylindrical, quartz vapour cell containing $^{85}$Rb atoms.
Quartz is absorptive at 12.2~$\mu$m, and thus the 12.2~$\mu$m BBR inside the cell should be in thermal equilibrium with the cell walls.
In this case, the atoms measure the average temperature of the walls of the vapour cell.
Two photon-counting photomultiplier tubes (PMTs) collect the fluorescence from the curved sides of the vapour cell.
Filter wheels installed between the PMTs and the vapour cell contain filters that individually isolate the 630~nm, 741~nm, and 762~nm fluorescence.
The signal recorded by the PMTs with a filter chosen to detect    wavelength of interest, $\lambda_{ij}$, between state $i$ and $j$ given the laser exciting state $k$ is then given by
\begin{equation}
     S^{(k)}_{\lambda_{ij}}  = N \Gamma_{ij} \eta_{\lambda_{ij}} n_i,
 \end{equation}
where, $N$ is the total number of atoms, and $\eta_{\lambda_{ij}}$ is the total detection efficiency at wavelength $\lambda_{ij}$.
The signal is measured in steady-state, such that $\dot{n}_i=0$.

For insight into the values of $n_i$, let us consider just the 5S, 5P, 5D, 7P, and 6D states and ignore degeneracy, fine structure, and hyperfine structure.
We find the steady-state populations of this simplified CoBRAS system by solving \eqref{eq:rate_equations} with the left-hand side set to zero:
\begin{equation}
    \begin{bmatrix} 0 \\ 0 \\ 0 \\ 0 \\ 0 \end{bmatrix} = 
    \begin{bmatrix}
    -\Omega_{\rm L} & +\Gamma_{5{\rm P}} & \Gamma_{\rm 5D} - \Gamma_{5{\rm D},5{\rm P}} & \Gamma_{7{\rm P}}-\Gamma_{7{\rm P},5{\rm D}}+\Omega_{\rm L} & \Gamma_{6{\rm D}}-\Gamma_{6{\rm D},5{\rm P}}\\
    0 & -\Gamma_{5{\rm P}} & +\Gamma_{5{\rm D}, 5{\rm P}} & 0 & \Gamma_{6{\rm D}, 5{\rm P}}\\
    0 & 0 & -\Gamma_{5{\rm D}} & +\Gamma_{7{\rm P},5{\rm D}} & 0 \\
   \Omega_{\rm L} & 0 & 0 & -\Gamma_{7{\rm P}} - \Omega_{\rm L} - \Omega_{7{\rm P}, 6{\rm D}} & \Omega_{7{\rm P}, 6{\rm D}} \\
   0 & 0 & 0 & \Omega_{7{\rm P}, 6{\rm D}} & -\Gamma_{6{\rm D}}-\Omega_{7{\rm P}, 6{\rm D}} \\
    \end{bmatrix}
    \begin{bmatrix} n_{5{\rm S}} \\ n_{5{\rm P}} \\ n_{5{\rm D}} \\ n_{7{\rm P}} \\ n_{6{\rm D}}
    \end{bmatrix}\label{eq:cobras_rate_eqn}\,,
\end{equation}
where $\Omega_{\rm L}$ is the laser-induced pumping rate from the $5{\rm S}$ to the $7{\rm P}$.
Because there are possible decay channels for an atom excited to the $5{\rm D}$ and $6{\rm D}$ states that do not go through $5{\rm P}$, we approximate these total rates as the differences $-\Gamma_{\rm 5D}+\Gamma_{\rm 5D,5P}$ and $-\Gamma_{\rm 6D}+\Gamma_{\rm 6D,5P}$, respectively.
The solutions for $n_{5{\rm D}}$ and $n_{6{\rm D}}$ yield 
\begin{eqnarray}
    \frac{n_{6{\rm D}}}{n_{5 {\rm D}}} = \frac{\Gamma_{5{\rm D}} \Omega_{7{\rm P},6{\rm D}}}{\Gamma_{7{\rm P},5{\rm D}} (\Gamma_{6{\rm D}}+\Omega_{7{\rm P},6{\rm D}})}\,,
\end{eqnarray}
which is independent of $\Omega_{\rm L}$, the laser pumping rate. 
If we further form the ratio
\begin{equation}
\label{eq:r_630_762}
r^{(7{\rm P})}_{630,762}=\frac{S^{(7{\rm P})}_{630}}{S^{(7{\rm P})}_{762}}=\frac{\eta_{630}}{\eta_{762}}\frac{\Gamma_{6{\rm D},5 {\rm P}}}{\Gamma_{5{\rm D},5{\rm P}}}\frac{\Gamma_{5{\rm D}} \Omega_{7{\rm P},6{\rm D}}}{\Gamma_{7{\rm P},5{\rm D}} (\Gamma_{6{\rm D}}+\Omega_{7{\rm P},6{\rm D}})}\,,
\end{equation}
we observe that it is independent of both $\Omega_{\rm L}$ and the number of atoms $N$.
Experimentally, this cancellation is quite useful as it effectively mitigates two common noise sources: laser intensity fluctuations and changes in the atom number.
Temperature dependence enters through $\Omega_{7{\rm P},6{\rm D}}$, and because $\Omega_{7{\rm P},6{\rm D}}\ll\Gamma_{6{\rm D}}$ for $T\sim300$~K, $r_{630,762}\propto U_\omega(\omega_{7{\rm P},6{\rm D}} \approx 2\pi\times24.6 \rm{THz})$.

While most parameters in \eqref{eq:r_630_762} are determined by the atomic physics, there is one parameter that is not: $\eta_{630}/\eta_{762}$.
One could determine this ratio of detection efficiencies $\eta_{630}/\eta_{762}$ by exciting to a higher-lying state like $8{\rm P}$, which would then spontaneously decay to both $5{\rm D}$ and $6{\rm D}$.
In this case, our simplified rate equations become
\begin{equation}
    \begin{bmatrix} 0 \\ 0 \\ 0 \\ 0 \\ 0 \end{bmatrix} = 
    \begin{bmatrix}
    -\Omega_{\rm L} & +\Gamma_{5{\rm P}} & \Gamma_{\rm 5D} - \Gamma_{5{\rm D},5{\rm P}} & \Gamma_{6{\rm D}}-\Gamma_{6D{\rm P},5{\rm P}} & \Gamma_{8{\rm P}}-\Gamma_{8{\rm P},6{\rm D}}-\Gamma_{8{\rm P},5{\rm D}}+\Omega_{\rm L}\\
    0 & -\Gamma_{5{\rm P}} & +\Gamma_{5{\rm D}, 5{\rm P}} & +\Gamma_{6{\rm D}, 5{\rm P}} & 0 \\
    0 & 0 & -\Gamma_{5{\rm D}} & 0 & \Gamma_{8{\rm P},5{\rm D}} \\
   0 & 0 & 0 & -\Gamma_{6{\rm D}} & \Gamma_{\rm 8P, 6D} \\
   \Omega_{\rm L} & 0 & 0 & 0 & -\Gamma_{8{\rm P}}-\Omega_{\rm L} \\
    \end{bmatrix}
    \begin{bmatrix} n_{5{\rm S}} \\ n_{5{\rm P}} \\ n_{5{\rm D}} \\ n_{6{\rm D}} \\ n_{8{\rm P}}
    \end{bmatrix}\label{eq:cobras_rate_eqn_systematic}\,,
\end{equation}
The corresponding ratio 
\begin{equation}
\label{eq:r_630_762_8P}
r^{(8{\rm P})}_{630,762}=\frac{S^{(8{\rm P})}_{630}}{S^{(8{\rm P})}_{762}}=\frac{\eta_{630}}{\eta_{762}}\frac{\Gamma_{6{\rm D},5 {\rm P}}}{\Gamma_{5{\rm D},5{\rm P}}}\frac{\Gamma_{8{\rm P},6{\rm D}}}{\Gamma_{6{\rm D}}}\frac{\Gamma_{5{\rm D}}}{\Gamma_{8{\rm P},5{\rm D}}}\,,
\end{equation}
is independent of $T$.
Thus, \eqref{eq:r_630_762_8P} can be used to solve for $\eta_{630}/\eta_{762}$ in terms of a measured quantity, $r^{(8{\rm P})}_{630,762}$, and several atomic physics parameters.
This independent measurement of $\eta_{630}/\eta_{762}$ would allow measurement of temperature through \eqref{eq:r_630_762} to be a {\it primary} temperature measurement, as it would not be traceable to a measurement of like kind.
However, this requires one additional laser.

Instead, we chose to determine $\eta_{630}/\eta_{762}$ by calibration, {\it i.e.}, fitting to a known temperature.
Figure~\ref{fig:cobras_experiment}(b) shows the measured ratio $r_{630,740}$.
We compute an expected ratio similar to \eqref{eq:r_630_762}, except we include all states with energy below $8{\rm S}$ in the rate equation model, and account for fine structure and the degeneracy due to hyperfine structure.
The one unknown parameter, $\eta_{630}/\eta_{762}$, is extracted from a fit to the data where $T>300$~K.
Temperature is measured using three industrial PRTs positioned inside a thermally insulating box of polystyrene foam that envelopes the vapour cell.
Temperature inside the box is stabilized and varied using a thermo-electric (Peltier) element.

The data and theoretical prediction follow each other quite closely for $T>300$~K with the single tunable fit parameter of  $\eta_{630}/\eta_{762}$.
For temperatures $T<300$~K, the Peltier element was run in cooling mode and this appears to have caused one or more potential problems including changing the thermal gradients inside the box and introducing water vapour condensation.
The root-mean-squared residuals are $<0.04~\%$ for $T>300$~K, corresponding to a temperature uncertainty of $u(T)\approx 0.13$~K.
This precision is the result of only 34~s of averaging, corresponding to approximately 0.75~K/$\sqrt{{\rm Hz}}$.

Here, the accuracy is set by the accuracy of the calibration itself--how well are the thermometers and thermal gradients known inside the box and at the surface of the cell--rather than the atomic physics parameters.
This perhaps surprising result is due to the fact that the shape of the theoretical prediction is not determined by the dipole matrix elements but by $U_\omega(\omega)$.

We also examined a temperature-independent ratio to check for systematics.
In particular, the ratio $r^{(7{\rm P})}_{741,762}$ results entirely from spontaneous emission.
(The simplified rate equation model is identical to \eqref{eq:cobras_rate_eqn_systematic} with the replacements $8{\rm P}\rightarrow 7{\rm P}$ and $6{\rm DS}\rightarrow 7{\rm S}$.)
Experimentally, as shown in Fig.~\ref{fig:cobras_experiment}, this ratio is temperature independent.
In a deployed CoBRAS, this ratio is potentially useful to measure potential systematic drifts in detection efficiencies.

In a potentially primary measurement, the theoretical prediction would be limited by the atomic physics parameters.
Fig.~\ref{fig:cobras_experiment}(b-c) also shows the potential range of values due to the uncertainty in the atomic dipole matrix elements, which is about 1~\% for $r_{630,740}$ and 10~\% for $r_{760,740}$.
These ranges give a sense for the expected uncertainty of such a primary CoBRAS thermometer.

\section{Conclusion}
Here, we have discussed the idea of building a thermometer based on the blackbody-induced transition rate, $\Omega_{ij}$, between two internal quantum states of an atom or molecule $i$ and $j$.
Presented in~\eqref{eq:transition_rate}, $\Omega_{ij}$ contains only fundamental constants and immutable constants of atomic and molecular physics, thus opening up the prospects for building a primary thermometer, in that it is not traceable to measurement of like kind.
Experimentally, measurement of $\Omega_{ij}$ requires measurement of some internal dynamics of an atomic or molecular system.
These full dynamics are only captured within a system of rate equations, introduced in Sec.~\ref{sec:rate_eqns}, of which $\Omega_{ij}$ is only a part.

We further discussed two experiments that attempted to use $\Omega_{ij}$ as a thermometer: the cold atom thermometer (CAT), presented in Sec.~\ref{sec:rydberg_thermometry} and the compact blackbody radiation atomic sensor (CoBRAS), presented in Sec.~\ref{sec:fluorescence_ratio}.
The CAT uses laser-cooled $^{85}$Rb Rydberg atoms to probe the BBR  component at 130~GHz (2.3~mm).
This experiment measures ratiometrically the relative populations in the initially excited 32S state to the subsequently blackbody-radiation populated 32P state, and extracts, through {\it ab initio} atomic theory, the temperature $T$.
This primary measurement is currently accurate at about 1~\% level, and further improvements could make it competitive with other techniques at the 0.1~\% level.
On the other hand, the CoBRAS uses a warm $^{85}$Rb vapour to probe the BBR component at 12.2~$\mu$m (24.5~THz).
This measurement uses fluorescence ratios from spontaeous decays of the BBR populated, and thus temperature-dependent, 6D state to the spontaneous-emission populated, and thus temperature-independent, 7S state.
While currently requiring calibration of one unknown experimental parameter, the ratio of two detection efficiencies, it has a statistical precision of $u(T)\approx 0.13$~K with only 34~s of averaging.

Both the CAT and the CoBRAS have already begun to reach the limitations of atomic theory.
Thus, a natural question arises: do the atoms measure temperature, or does accurate knowledge of the thermal environment now allow for better measurement of atomic properties.
Indeed, BBR represents one of the most accurately known sources of electromagnetic radiation, which is why it is used ubiquitously in radiometry for calibration.
The total relative uncertainty of BBR can be $<10^{-4}$, limited by knowledge of the emissivity and temperature.
This is two orders of magnitude better than typical knowledge of $\left|\left<i|d|j\right>\right|^2$.
Thus, it would appear that we are poised to dramatically increase our knowledge of atomic physics parameters using BBR. 

While that might seem antithetical to the goal of creating an atom- or molecule-based radiation thermometer, such a thermometer would still have the advantage immutability properties.
Specifically, although the accepted values of $\left|\left<i|d|j\right>\right|^2$ will be traceable to the kelvin and therefore not strictly primary, the constancy of this value will still allow for calibration-free radiation thermometers.
Such thermometers might be dubbed ``semi-primary.''
Considering that radiometer drift is a large problem in remote sensing and other applications, such a constant, stable, semi-primary thermometer would still represent a great advantage.\vskip6pt

\begin{acknowledgments}
We thank Eric Shirley and Sai Paladugu for useful discussions and Daniel Barker and Dixith Manchaiah for a thorough reading of the manuscript.
This work was supported by the National Institute of Standards and Technology.

Data presented within the text has been previously published in Refs.~\cite{schlossberger_primary_2025, mantia_compact_2024}.
The authors declare no competing interests.
\end{acknowledgments}    
\bibliographystyle{RS}
\bibliography{references, other}

\end{document}